# Consistency of Commitments in Social Web Services

Marzieh Adelnia
*Department of Computer Engineering*
*Sheikh Bahaee University*
*Isfahan, Iran*
adelniya_m@yahoo.com

Mohammad Reza Khayyambashi
*Department of Computer Engineering*
*University of Isfahan*
*Isfahan, Iran*
m.r.khayyambashi@eng.ui.ac.ir

*Abstract* - Web Service is one of the most important information sharing technologies on the web and one of the example of service oriented processing. To guarantee accurate execution of web services operations, they must be accountable with regulations of the social networks in which they sign up. This operations implement using controls called "Commitment". This paper studies commitments, then has an overview on existing researches, web service execution method using commitments and information sharing methods between web services based on commitments and social networks. A key challenge in this technique is consistency ensuring in execution time. The aim of this study is presenting an algorithm for consistency ensuring between commitments. An application designed for proving correctness of algorithm.

*Index Terms – Commitments, Web Service, Consistency, Social Networks, Social Web Service.*

## I. INTRODUCTION

Web Service is one of the important information for sharing technology on the web and one of the examples of service oriented processing. According to the W3C, a Web service "*is a software application identified by a URI, whose interfaces and binding are capable of being defined, described, and discovered by XML artifacts and supports direct interactions with other software applications using XML based messages via Internet-based applications*".

In recent years, there has been an increasing interest in social networks and Web Services. Regarding the growth of social networks and a tendency for joining those, web services can be looked from a new perspective named social computing. A prime example for those processes is application based on WEB2 like social networks and blogs. Social computing can analyze how to composite and share web services, keep information security and have fault tolerance [1].

On the other hand, service-oriented computing is the development of applications based on the theory that says: "*I offer services that somebody else may need*" and "*I require services that somebody else may offer*" [2].

With merging service oriented computing and social computing, social web services can be produced that are more complicated from regular web services. To guarantee accurate execution of social web services operations, they must be accountable and adjustable with regulations of the social networks in which they sign up [1]. These operations implement using controls called "Commitment". In other words, transactions between web services components and social networks lead to creation, management and using of commitments [3].

## II. BACKGROUND

This section provides an overview of social web services and regular web services consisted of commitments.

### A. Overview on Social Web Services.

The synergy between social computing and service oriented computing has eventuated into social web services. Existing research focuses on adopting web services to social networks.

Maaradji proposed a social constructor named "SoCo" to suggest and help users for next their operations (like selecting specific a web service). So users may like to perform an operation that their friends have done in social networks [4].

Maamar purposed an approach for weaving social networks operation using web services. The result of his researches lead to creating social web services [5].

In the other research, Maamar et al. categorized social networks to three group including [6]:

- Collaboration social networks. "By emerging their respective functionalities, social Web services have the capacity to work together and response to complex user requests. In fact, a social Web service manages its own network of collaborators".
- Substitution social networks. Although social Web services compete against each other, they can still help each other when they fail as long as they offer similar functionalities.
- Competition social networks. Social Web services compete against each other when they offer similar functionalities. Their non-functional properties differentiate them when users' non-functional requirements must be satisfied Overview on Commitments.

### B. Overview on commitments

First time, Fornara and Colombetti defined a general Formula for commitments. They used commitments for speech evaluation [7].

Bentahar et al. proposed a new persuasion dialogue game for agent communication. They modeled dialogue game by a framework based on social commitments and arguments, Called Commitment and Argument Network (CAN). This framework allows to model communication dynamics in levels of activities that agents apply to commitments and in levels of argumentation relations. This dialogue game is specified by indicating its entry conditions, its dynamics and its exit conditions. They proposed a set of algorithms for the implementation of the persuasion protocol and discuss their termination, complexity and correctness [8].

Singh et al. are the first of few who advocated for examining Service-Oriented Architecture (SOA) principles from a commitment perspective. As regarding Existing service-oriented architectures are formulated in terms of low level abstractions far removed from business services. In CSOA[1], the components are business services and the connectors are patterns, modeled as commitments, which support key elements of service engagements [2].

El-Menshawy et al. showed that current approaches fail to capture the meaning of interactions that arise in real-life business scenarios and proofed commitments increase flexibility and intuitively in protocols. He presented an exploder definition for commitments for using in a larger level. In his definition, a new grammar named *CTL* and terms like $SC^P$ for unconditional commitments and $SC^C$ for conditional commitments was added. In fact, *CTL* is a logical tree and commitments are the nodes of tree that organize in tree base on logical regulation in transaction execution time [9].

Narendra represented a contract as a collection of the participants' commitments toward each other. The interactions that take place in a contract are understood in terms of how they operate on the participants' commitments. The operations on a commitment cause its state to change according to a life cycle [10].

Grosof and Poon defined a rule-based approach for e-contracting. In their approach a contract is a set of activities that can be decomposed into sub-activities. The terms of contracts uses a set of commitments for execute operation by agents. Algorithm uses a coordination method to manage agent activities [11].

*C. Overview on Consistency in Web Services*

Choi et al. presented a mechanism to insure consistency for web services transactions. This mechanism recognize inconsistent states of transactions and replace them with consistent states. Mechanism operation is designed by a waiting graph of web services transactions and a coordinator that check waiting graph. If coordinator is certified about deadlock lack, allow transaction to execute. Also if deadlock occurred, coordinator recognized a safe state by using waiting graph and replace it instead deadlock state. Based on this mechanism, web service transaction dependency management protocol named WTPD is designed and presented [12].

Reiko et al. suggested an algorithm to guarantee consistency of web services. It receives activity diagram of web service and translate it into CSP to be analyzed for deadlock freedom and protocol consistency [13].

Shan-liang designs a modeled for transaction processing coordination model based on BPEL. In this model a coordinator is used for web services transaction weaving and if deadlock occurred coordinator rollback web services activities [14].

Greenfield et al. developed a protocol for dynamic consistency checking that can be run at the termination of a service-based application [15].

### III. COMMITMENT DEFINITION

This section provides a definition of commitments and the types defined for it.

*A. Types*

Two types of commitments are identified [1]:
1. Social Commitments: guarantee the proper use of the social networks in which the social Web services sign up.
2. Business Commitments: guarantee the proper development of composite Web services in response to users' requests.

*B. Structure*

Maamar et al. define a formula for Social Commitments based on Fornara's formula. Fornara and Colombetti note that *"...intuitively a social commitment is made by an agent (the debtor) to another agent (the creditor), that some fact holds or some action will be carried out (the content)"*. In addition to this formula, Maamar considers a list of responsibility for social web services. In fact a commitment structure can be designed as: $C_{Resp_i}$ (debtor, creditor, content [|condition]). Condition parameter is optional [3].

Also they define business commitment similar to social commitment with the difference that in these commitments a social web service assigns to a debtor and compositions assign to a creditor [1].

Social commitments defined by Maamar listed as follows [1]:
1. *"Resp1. Collecting any detail (d) in a social network would require indicating the purpose (p) of this collection to this detail's owner (o), represented as Permission(Collect(d, o, valid(p))). Collect is the action, d is for instance a non-functional property like response time, o is the owner of d for instance social Web service, p is the rationale of collecting d, and valid is a function that checks p.*
2. *Resp2. Posting any detail (d) on a social network should be correct. It can be represented as Obligation(Post(d, true)). Post is the action of web service and true is the veracity of d.*
3. *Resp3. Collecting any detail (d) from a social network should not be tampered after information*

---
[1] Commitment-based SOA

*collection. This responsibility can be represented as Obligation(not-Tamper(d, o, collection(d))). not-Tamper is the operation and collection is a function that checks if collecting d is approved in compliance with Resp1.*
4. *Resp4. Signing off from a social network would require the completion of all the pending assignments (ass). It can be represented as Permission(Signoff(status(ass))). Sign-off is the action and status is a function that assesses the progress (e.g., ongoing, complete, and failed) of ass.*
5. *Resp5. Revealing any public detail (d) to the non-members (not(m)) of a social network should not be authorized indefinitely, represented as Obligation(not-Reveal(d, o, m, collection(d))). not-Reveal is the action, m corresponds to the non-members of a social network, and collection is a function that checks if collecting d is approved in compliance with Resp1."*

## IV. CONSISTENCY CHECKING ALGORITHM

Maamar et al. formulate Social Networks operation based on Commitment concept. They simulate Web Service actions using Commitments. They define 5 commitment for social networks.

Based on the effect of commitment on social network information, commitments can be categorized to followed groups:
1. Reader Commitments: this category of commitments doesn't change the information of database and usually act as an information collector for other social networks or purpose checker in social web services. Note that the purpose of social web services that use reader commitments must be valid. Also privacy must be protected.
2. Writer Commitments: unlike reader commitments, this category can change the information of database and social networks. They share Information and Post activity on other social networks. So writer commitments are more effective than reader commitments in social web service transactions. Like reader commitments, in writer commitments privacy of information must be controlled.

As regards commitments implement the action of social web services and both reader and writer commitments may act on social networks concurrently, a major problem that must be considered is consistency ensuring of commitments in social web services.

For achieve this consistency, commitment must have priority property. Because of:

1. Sometimes, if two or several social web services are ready for execution, it is important what action is executed first.
2. Private privacy levels have higher priority than public privacy levels.

Assigning priority to commitments must be accomplished carefully.

To ensure consistency of commitments, three concept are considered as follows:

1. Friend: the commitment are friendly if they are reader commitments. So they are consistent in all states and database is in the safe state.
2. Family: commitments are family if they are writer. In fact they effect on the database and information state.
3. Strange: if commitments neither friend nor family are strange. In this state, commitments may be reader or writer.

Since writer commitments have effect on database and information, so if active commitments of social web services be family or strange with other, conflict may occurred. In this case consistency must be guaranteed and if deadlock happens it would be removed and system need to be recovered.

When a web service sign up in a social network, it is recognized by authority component, if it is accepted, responsibility is assigned to web service and its commitments will be created. This time, consistency checking between active commitments is critical and vital. To guarantee consistency, first current commitment condition is checked towards active commitment. Three conditions may occurred:

1. IsFriend: if current commitment and active commitment are friend, both can execute concurrent.
2. IsFamily: if current commitment and active commitment are family, current commitment would wait until active commitment execution is finished.
3. IsStrange: if current commitment and active commitment are strange, current commitment would wait until active commitment execution finished.

Sometimes several commitments are created concurrent in a time slice. In this state, commitments can be executed based on two policy:

1. FCFS: commitments service based on order input time. This policy is fairness.
2. Priority: commitments service by priority.

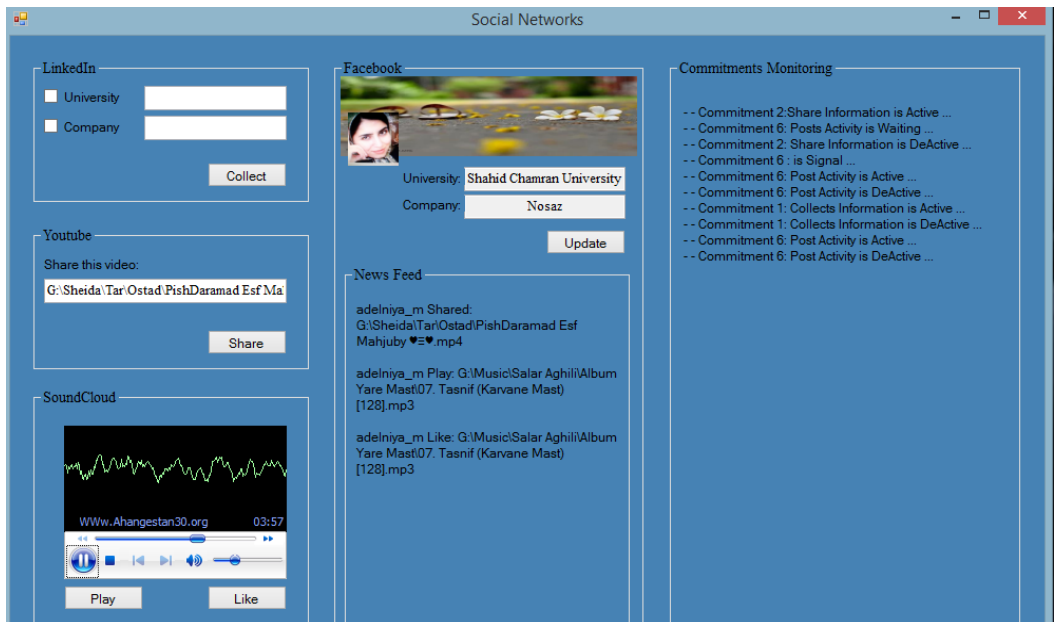

Figure 1. Algorithm Implementation

## V. IMPLEMENTATION

An application designed for suggestion algorithm. The commitments architecture that is implemented by Maamar is used as base implementation. In suggestion algorithm, priority, IsReader and IsWriter property have been added to Commitments architecture. A queue is used for waited commitments.

In this application five section considered as follows:

- LinkedIn: this section simulates LinkedIn social network. For simplicity suppose that it collects some information from other social networks only.
- YouTube: this section simulates YouTube social network operation like video sharing.
- SoundCloud: it is a music social network. Some activity like play music, like and share can be done in it.
- Facebook: this section is central social network in this implementation that communicates with 3 other social networks.
- Commitments Monitoring: for monitoring state of commitments in any time, this section designed and shows the number and state of any commitments for all active social web services in a social network.

Experiments:

To carry out experiments, all possible states may occurred in execution of social web services considered and checked.

First, a web service sign up in a social network, if it is authenticated by authority component, it would registered in social network and changed to a social web service. Then one or several responsibility assign to this social web service. Any responsibility has default commitment that act on user account and user information. For a user in social network, if no commitment is active, commitments of responsibility could be active and execute their operations. But if another commitment is active on this user account and information, consistency must be protected. Thus application checks the state of current commitment towards active commitment and decide commitments execute or wait.

## VI. CONCLUSION

This study set out to present an algorithm to ensure consistency of commitments in social web services. Also two commitments are considered and added to base commitments for optimizing. This study categorizes commitments into two groups contain reader and writer commitments. Algorithm is designed and described using base concept in social network like "Family" and "Friend". For designing algorithm three properties have been added to commitments structure contain reader, writer and priority. If commitments only collect information, called reader commitments and if they affect and change information and database, called writer commitment. Commitments may have three state into each other. They may be friend, family or strange based on their operations. Algorithm manages different states that may occurred in commitment execution of a social web service operation. An application is implemented for correctness checking.

## VII. REFERENCES

[1] Z. Maamar, N. Faci, K. Boukadi, Q. Z. Sheng and L. Yao, "Commitments to Regulate Social Web Services Operation," *IEEE Transaction on Service Computing,* vol. 7, no. 2, pp. 154-167, 2013.

[2] M. P. Singh, A. K. Chopra and N. Desai, "Commitment-Based Service-Oriented Architecture," *Computer,* vol. 42, no. 11, pp. 72-79, Nov 2009.

[3] Z. Maamar, N. Faci, M. Luck and S. Hachimi, "Specifying and Implementing Social Web Services Operation using Commitments," *SAC,* pp. 1955-1960, ACM 2012.

[4] H. H. J. D. a. N. C. A. Maaradji, "Towards a Social Network Based Approach for Services Composition," in *2010 IEEE International Conference on Communications (ICC'2010)*, 2010.

[5] Z. Maamar, H. Hacid and M. N. Huhns, "Why Web Services Need Social Networks," *IEEE Internet Computing,* vol. 15, no. 2, pp. 90-94, 2011.

[6] Z. Maamar, H. Yahyaoui, E. Lim and P. Thiran, "Social Engineering of Communities of Web Services," in *the 11th Annual International Symposium on Applications and the Internet (SAINT'2011)*, Munich, Germany, 2011.

[7] N. Fornara and M. Colombetti, "Operational Specification of a Commitment-Based Agent Communication Language," in *AAMAS'02*, Bologna, Italy, July 15-19, 2002.

[8] J. Bentahar, B. Moulin and B. Chaib-draa, "Specifying and Implementing a Persuasion Dialogue Game Using Commitments and Arguments," in *Argumentation in Multi-Agent System*, 2005, pp. 130-148.

[9] M. El-Menshawy, J. Bentahar and R. Dssouli, "Verifiable Semantic Model for Agent Interactions using Social Commitments," in *Languages, Methodologies and Developement Tools for Multi-Agents Systems*, 2010, pp. 128-152.

[10] N. C. Narendra, "Generating Correct Protocols from Contracts: A Commitment-based Approach," *SERVICES I,* pp. 407-414, IEEE Computer Society 2008.

[11] B. N. Grosof and T. C. Poon , "SweetDeal: Representing Agent Contracts With Exceptions using XML Rules, Ontologies, and Process Descriptions," in *12th international conference on*, Budapest, Hungary, 2003.

[12] S. Choi, H. Kim, H. Jang, J. Kim, S. M. Kim, J. Song and Y.-J. Lee, "A framework for ensuring consistency of Web Services Transactions," *Information and Software Technology,* no. 50, p. 684–696, 2008.

[13] H. Reiko, V. Hendrik, J. KU¨STER and S. THO¨NE, "Towards Consistency of Web Service Architectures," University of Paderborn, Paderborn, Germany, 2004.

[14] P. Shan-liang, L. Ya-Li and L. Wen-juan, "A framework for ensuring consistency of Web Services Transactions based on WS-BPEL," *Modern Education and Computer Science,* no. 4, pp. 47-54, 2011.

[15] P. Greenfield, D. Kuo, S. Nepal and A. Fekete, "Consistency for Web Services Applications," *31st international conference on Very large data bases(VLDB),* pp. 1199-1203, 2005.